\documentclass{article} 
 \usepackage{mystyle-new}
 \usepackage{authblk}
 \usepackage{epsfig,amsmath}
\usepackage{graphicx}
\usepackage{hyperref}
\usepackage{authblk}
\usepackage{orcidlink}
\usepackage{amsmath,amssymb}
\usepackage{abstract}
\usepackage{lineno}
\emergencystretch=\maxdimen
\title{Non-Perturbative Contributions to Low Transverse Momentum Drell-Yan Pair Production Using the Parton Branching Method
\date {}
\thanks{Presented at XIII International Conference on New Frontiers in Physics 2024 (ICNFP 2024)}} 
\author{Nata\v sa~Rai\v cevi\' c on behalf of the CASCADE Group}
\affil{University of Montenegro, Faculty of Science and Mathematics\\
 Podgorica, Montenegro}
\begin{document}
\maketitle
\begin{abstract}
\noindent
The non-perturbative processes, the internal transverse motion of partons inside hadrons, which gives rise to their intrinsic transverse momentum (intrinsic 
$k_{\rm T}$), and multiple soft gluon emissions that need to be resummed, are dominant contributors to the low transverse momentum of the Drell-Yan (DY) pair cross section. Therefore, this part of the DY spectrum serves as a powerful tool for a better understanding of such processes, which is the focus of the study presented here. The study is conducted using the Parton Branching Method, which describes Transverse Momentum Dependent (TMD) Parton Densitity Functions (PDF) and provides a very precise description of DY pair transverse momentum distributions across a wide range of collision energies and pair invariant masses.
In contrast to the energy dependence of intrinsic $k_{\rm T}$ observed in shower-based Monte Carlo event generators, the CASCADE3 event generator—based on the Parton Branching Method —has provided an intrinsic-$k_{\rm T}$ distribution that is independent of the center-of-mass energy. 
Further studies conducted within the Parton Branching Method have sought to understand the origin of this energy dependence, indicating that the dependence is mainly a consequence of the interplay between two main processes: internal transverse motion and soft gluon emission. The latter has been reduced in shower-based event generators, primarily due to the non-perturbative Sudakov form factor, which is often neglected.
Since the Sudakov form factor depends on the evolution scale, this paper explores this dependence through the interplay of the two processes and attempts to explain it. Additionally, since QED final state radiation affects the profile of the DY pair transverse momentum distribution, we investigate its impact in both the high and low DY pair invariant mass regions.
\end{abstract}

\newpage

\label{sec:intro}
\section{Introduction}

Recently, considerable attention has been focused on the impact, determination, and explanation of the trend in the intrinsic transverse momentum of partons inside the initial hadrons~\cite{ ktpaper, cms_gen, pbescaling, gluon, natasa_hs, natasa_qcd}. These studies analyzed the transverse momentum distribution of Drell-Yan (DY) pairs, which is considered the most convenient tool for determining intrinsic-$k_{\rm T}$ because it has a significant impact on the lowest transverse momenta of the pairs. The primary goal of this research was to provide the most precise determination of the theoretical predictions in order to offer high-precision expectations for the Standard Model, which are necessary for potentially new physics and discoveries.
\\

The "problem" that led to the intensive study of this subject was highlighted by research~\cite{ktpaper} conducted with the CASCADE3~\cite{cascade} event generator, which is based on the Parton Branching Method (PB)~\cite{h1,h2,pb} together with an NLO calculation for the hard process from  ${\rm {MADGRAPH5{\_}AMC@NLO}}$~\cite{madgraph}. The final state parton shower, as well as the QED radiation, was generated using PYTHIA~\cite{pythia}. In this study, the intrinsic-$k_{\rm T}$ distribution was determined, and it was shown that the internal transverse motion is not affected by the initial hadron's energy or the mass of the created DY pair. This finding contrasts with results obtained many years ago using standard Monte Carlo event generators PYTHIA and HERWIG~\cite{shmc1, shmc2}, which were recently confirmed by the CMS collaboration~\cite{cms_gen}.  
\\

Intrinsic-$k_{\rm T}$ is introduced at the starting evolution scale, $\mu_0$, through integrated PDFs in standard Monte Carlo event generators or through TMDs in the PB Method and CASCADE3. This is achieved by multiplying the density function at $\mu_0$ by a Gaussian distribution with zero mean and width $\sigma$ which describes the intrinsic-$k_{\rm T}$ distribution. Therefore, at the starting scale, there is a component related to the transverse motion, represented by the exponent $e^{\frac {-k_T} {\sigma^2}}$.  The width $\sigma$ is related to the parameter $q_s$ used in the PB Method by the relationship $\sigma = \sqrt 2 q_s$.  
\\

Recently, the CMS Collaboration obtained results on the dependence of the intrinsic-$k_{\rm T}$ width  on the center-of-mass collision energy~\cite{cms_gen} for PYTHIA (CP5, CP4, and CP3 tunes) and HERWIG (CH3 and CH2 tunes). These results clearly show the energy dependence of the intrinsic-$k_{\rm T}$ width, where the width increases significantly with the collision energy. This result contrasts with the findings obtained using the CASCADE3 event generator, as published in~\cite{ktpaper}, which shows no significant, or only a very mild, dependence of the intrinsic-$k_{\rm T}$ width on the center-of-mass collision energy.
 The main difference between shower-based event generators and CASCADE3 lies in their treatment of soft parton emissions. While shower-based event generators impose a minimum transverse momentum that a parton must have, CASCADE3, by its design, includes all soft gluon contributions.
\\

To understand the origin of the energy dependence and why it is absent in the case of the CASCADE3 event generator, a parton shower generator was mimicked within the PB Method. This was achieved by applying a cutoff on the transverse momentum  of the parton emitted in each branching ($q_{\rm T}$), specifically $q_{\rm T} > q_0$ as done in~\cite{pbescaling}.
It was observed that as $q_0$ increases—indicating more excluded soft gluon contributions—the dependence of the intrinsic $k_{\rm T}$ width on 
$\sqrt s$ becomes steeper. Therefore, it was concluded that the origin of the  intrinsic $k_{\rm T}$ width energy dependence arises from the reduction of real emissions, as well as from the non-perturbative part of the Sudakov form factor.
\\

The study presented in this paper is a continuation of the research highlighted above and focuses on the soft gluon contributions that interplay  with the internal transverse motion in the region of very low transverse momentum of DY pairs. A detailed analysis is performed on the scale dependence of the non-perturbative Sudakov form factor,  which, as concluded in the previous research, influences the energy dependence of the intrinsic $k_{\rm T}$ width due to the interplay of the two processes.  
\\

The paper is organized as follows. Section~\ref{sec:dy} discusses the influence of the intrinsic-$k_{\rm T}$ width on the shape of the DY pair transverse momentum distribution and examines the interplay between internal motion and soft gluon contribution at very low pair transverse momentum. 
Section~\ref{sec:nps} addresses the influence of soft gluon emission contributions on the intrinsic-$k_{\rm T}$ width, focusing particularly on the scale dependence of the Sudakov form factor.
This section presents new results on the dependence of the intrinsic-$k_{\rm T}$ width and the soft gluon contribution on the DY pair invariant mass, which is directly related to the scale in the Sudakov form factor. The potential influence of QED radiation on the results and conclusions related to the non-perturbative processes is discussed in Section~\ref{sec:qed}. The paper is concluded with Section~\ref{sec:concl}.

\section{The interpaly of the non-perturbative processes at low DY pair transverse momenta}
\label{sec:dy}

As pointed out above, the intrinsic-$k_{\rm T}$  energy dependence in the PB Method is introduced by applying a cut on the minimum parton transverse momentum emitted in the branching, which excludes a certain fraction of  soft contributions and that, as shown in~\cite{pbescaling},  as this cut, $q_0$, decreases, the slope of the energy dependence of the intrinsic-$k_{\rm T}$ width also decreases.  Only when this value is close to 0 GeV does the energy dependence disappear. Specifically, when $q_0 \simeq 0$~GeV, the $q_s$ width is related to pure intrinsic-$k_{\rm T}$. In other cases, it results from the interplay between internal motion and soft gluon emission; therefore $q_s$ increases with the $q_0$ cut.  

\begin{figure} [t!]
\includegraphics[width=.5\linewidth]{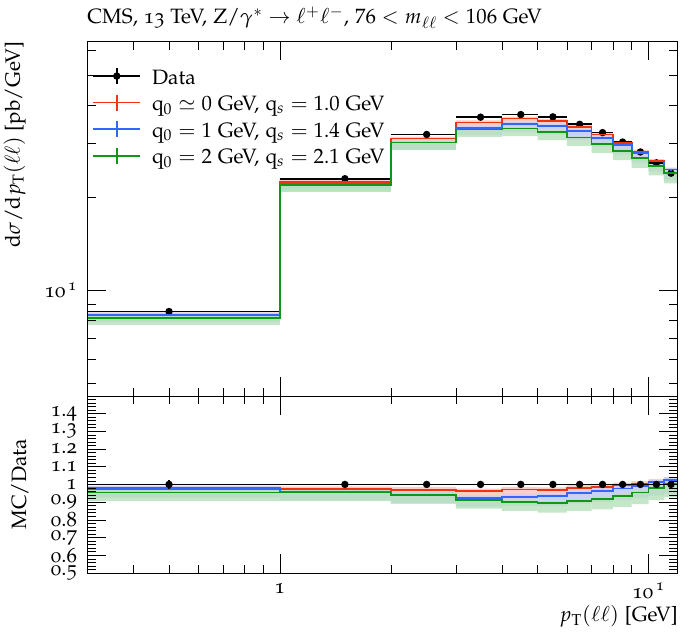}
\includegraphics[width=.5\linewidth]{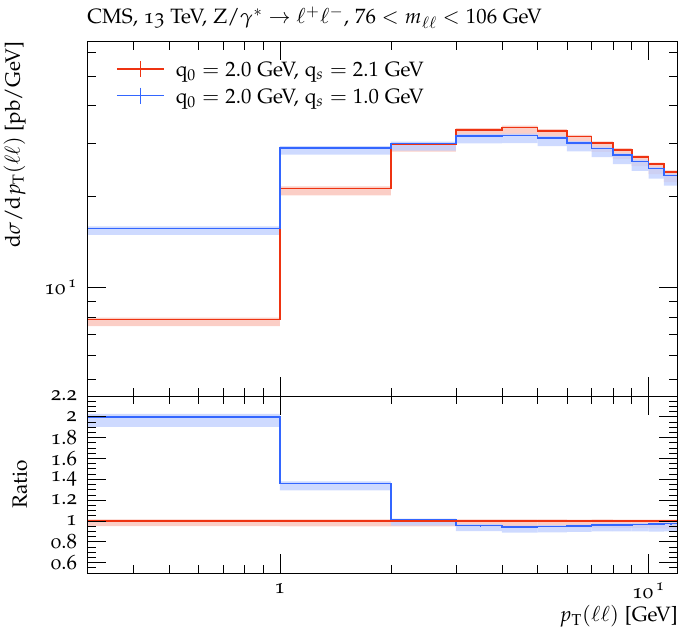}
\caption{(Left) Measurement of the cross section at $\sqrt s = 13$~TeV~\cite{cms_2022}  as a function of DY pair transverse momentum, compared with CASCADE3 predictions in the 
Z-peak region for three different values of $q_0$. (Right) 
CASCADE3 prediction of the cross section in the Z-peak region as a function of DY pair transverse momentum at $\sqrt s = 13$~TeV
obtained with  $q_0 = 2.0$~GeV using the optimal value of $q_s$ (red histogram) and the value of $q_s$ obtained for $q_0 \rightarrow 0$ (blue histogram). The bands show the scale uncertainty.}
\label{fig:xs_vs_q0}
\end{figure}

The optimal value of the intrinsic-$k_{\rm T}$ width was determined by minimizing of the  $\chi^2 (q_s)$ calculated from the difference between the measurements and the predictions obtained for different values of the $q_s$ width, as described in detail in~\cite{ktpaper, pbescaling}. The comparison of the measured cross section as a function of DY pair transverse momentum in the Z-peak region ($76 \le m_{DY}(ll) \le 105$~GeV) with the predictions from CASCADE3 event generator, using final distributions obtained with the Rivet tool~\cite{rivet} for three values of $q_0$ is shown in the left part of Figure~\ref{fig:xs_vs_q0}. 
This figure, along with the other figures in the text that compare data and prediction distributions, also shows the scale uncertainty through the bands obtained from the 7-point variation of the renormalization and factorization scales.
The values $q_s$  used in each prediction are the optimal ones obtained through the mentioned minimization procedure. The $q_s$ values are indicated in the figure. After adjusting the $q_s$ value, a slight difference remains in the transition region (5-10~$\%$) due to soft gluon contributions that were excluded by the $q_0$ cut. The value of $q_s$ which is obtained with $q_0 \simeq 0$~GeV can be treated as a pure intrinsic-$k_{\rm T}$ and as  shown in~\cite{ktpaper}, it exhibits no pronaunced  dependence on collision energy, $\sqrt s$, neither on DY pair invariant mass.. The determined value is $1.04 \pm 0.08$~GeV. It can be seen that the optimal value of $q_s$ for $q_0 = 2.0$~GeV is about twice as large as this one. 
\\

As indicated in~\cite{pbescaling,gluon}, the increase of the intrinsic-$k_{\rm T}$  with increase of $q_0$  is related to the Sudakov form factor $\Delta( \mu^2 , \mu^2_0 )$ which gives the probability of no radiation between the two values of the scales, $\mu_0$ and $\mu$:    

\begin{eqnarray}
\label{eq:divided_sud}
 \Delta( \mu^2 , \mu^2_0 ) & = &
 \exp \left(  -  \sum_b  
\int^{\mu^2}_{\mu^2_0} 
{{d {\bf q}^{\prime 2} } 
\over {\bf q}^{\prime 2} } 
 \int_0^{z_M} dz \  z 
\ P_{ba}^{(R)}\left(\alpha_s , 
 z \right) 
\right) 
\end{eqnarray}
\\
Here, $z$ is the fraction of the longitudinal momentum transferred at the branching and $P_{ba}^{(R)}$ is resovable splitting function which describes splitting of parton $b$ to parton $a$~\cite{h1,h2}. The transverse momentum of the parton emitted through this splitting in the branching is related to the branching variable ${\bf {q'}^2}$, according to the angular ordering~\cite{ao} as $q_{\rm \perp} = (1 - z) |{\bf q'}|$. 
$z_{\rm M}$ is a parameter introduced as $z_{\rm M} = 1 - \epsilon$ to bring $z$ as close as possible to 1 through a numerical approach by achieving $\epsilon \rightarrow 0$. 
\\

According to the angular ordering, there is a value of the parton transverse momentum emitted during the branching, denoted as $q_0$ , which leads to two regions: a perturbative region where $q_{\rm \perp} > q_0$  and a non-perturbative region where  $q_{\rm \perp} < q_0$, in which  $\alpha_s$ is frozen~\cite{pbescaling}. Therefore, it is possible to introduce an intermediate $z$-scale, referred to as dynamic-$z$ defined as:

\begin{equation}
z_{\rm {dyn}} = 1 - q_0/q'
\end{equation}
\\
By introducing $z_{\rm {dyn}} $, the integral in~\ref{eq:divided_sud} can be split in two parts as follows:

\begin{eqnarray}
\label{eq:divided_sud1}
 \Delta( \mu^2 , \mu^2_0 ) & = &
\exp \left(  -  \sum_b  
\int^{\mu^2}_{\mu^2_0} 
{{d {\bf q}^{\prime 2} } 
\over {\bf q}^{\prime 2} } 
 \int_0^{z_{dyn}} dz \  z 
\ P_{ba}^{(R)}\left(\alpha_s , 
 z \right) 
\right) \nonumber \\
& & 
 \times \exp \left(  -  \sum_b  
\int^{\mu^2}_{\mu^2_0} 
{{d {\bf q}^{\prime 2} } 
\over {\bf q}^{\prime 2} } 
 \int_{z_{dyn}}^{z_M} dz \  z 
\ P_{ba}^{(R)}\left(\alpha_s , 
 z \right) 
\right) \nonumber \\
& = &  \Delta^{(\text{P})}\left(\mu^2,\mu_0^2,q^2_0\right)  \cdot \Delta^{(\text{NP})}\left(\mu^2,\mu_0^2,q_0^2\right) \; .
\end{eqnarray}
\\
In the PB Method $z_{\rm M} \rightarrow 1$  while in shower-based event generators $z_{\rm M} = z_{\rm {dyn}}$.  This implies that when mimicking shower-based MC event generators,  the non-perturbative part of the Sudakov form factor, $\Delta^{(\text{NP})}\left(\mu^2,\mu_0^2,q_0^2\right)$, is neglected through the integral in $z$. 
\\

To ilustrate the effect of the excluded soft gluon contribution, we show in the right part of Figure~\ref{fig:xs_vs_q0} a comparison of two predictions obtained by CASCADE3 for $q_0 \simeq 2$~GeV: one with a pure intrinsic-$k_{\rm T}$ value, $q_s \simeq 1$~GeV, and another one with an optimal value $q_s = 2.1$~GeV. As shown in the left part of the figure, the latter prediction provides a good description of the data at the lowest DY pair transverse momenta. This figure demonstrates a factor of two difference between the predictions up to 1~GeV in the pair transverse momentum. This clearly indicates the contribution from soft gluons, which is excluded by the $q_0$ cut and primarily affects the lowest pair transverse momentum region, interplaying with the intrinsic-$k_{\rm T}$.

\section{The impact of the scale dependence of the Sudakov form factor on the $q_s$ width}
\label{sec:nps}

As can be seen from the formula~\ref{eq:divided_sud1}, the Sudakov form factor depends not only on  $z_{\rm {dyn}}$ but also on the evolution scale $\mu$.  Therefore the scale should impact   the intrinsic-$k_{\rm T}$ width $q_s$ through the interplay of the two processes. If the width, $q_s$ results solely from intrinsic motion within the hadron, it should not depend on the pair invariant mass. However, if it also accounts for the missing soft gluon contribution in the Sudakov form factor, then such dependence should emerge.  Since the scale is related to the DY pair invariant mass, we explore the dependence of $q_s$  on the invariant mass.  The interplay between the two processes and the impact of the Sudakov form factor on $q_s$ through its $\mu$ dependence  is studied in detail in this section. 

\begin{figure} [t!]
\centering
\includegraphics[width=.45\linewidth]{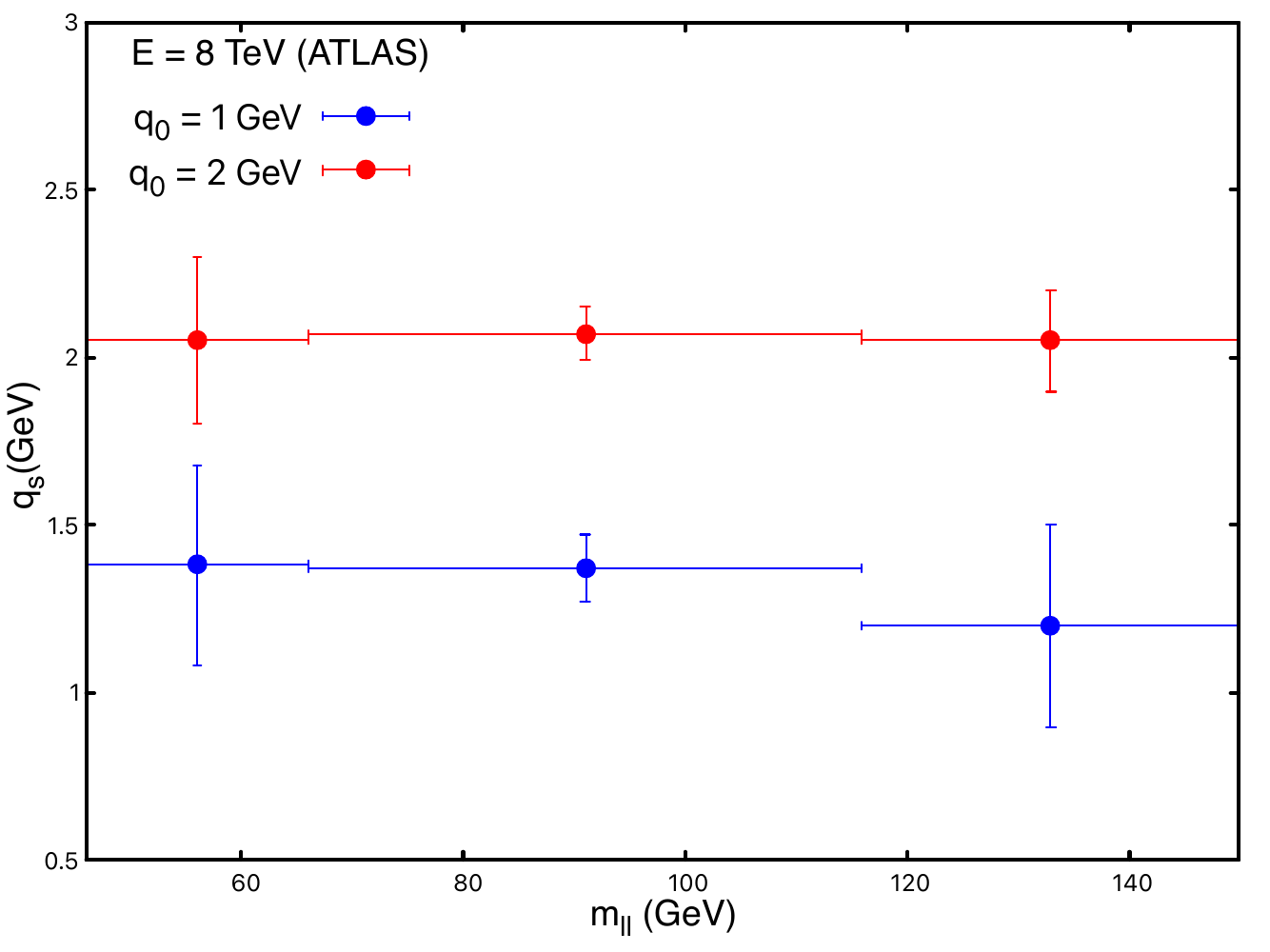}
\includegraphics[width=.45\linewidth]{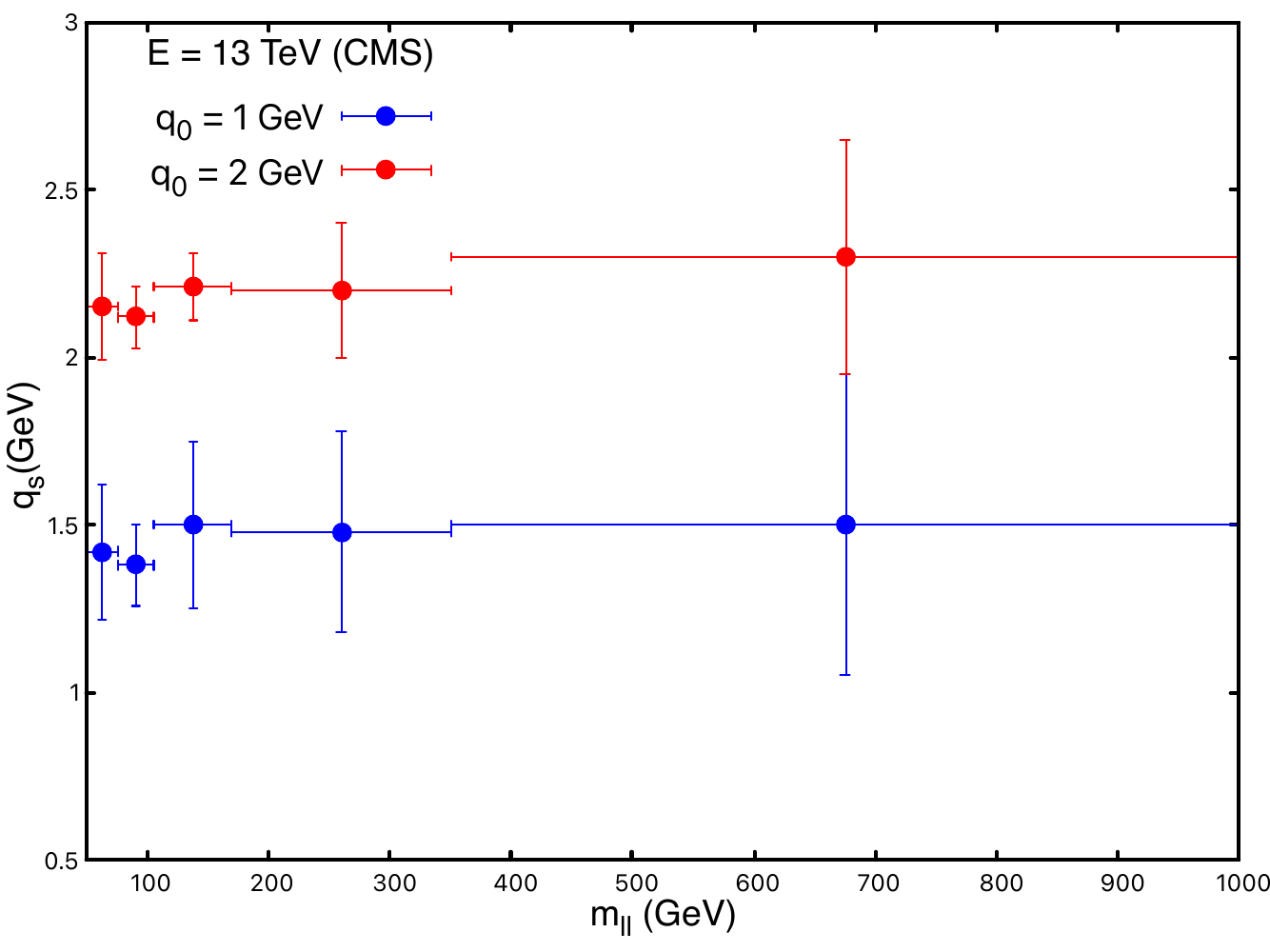}
\caption{Intrinsic $k_{\rm T}$ width as a function of DY pair invariant mass, obtained from the comparison of CASCADE 3 predictions with two values of $q_0$ and measurements obtained at $\sqrt s = 8$~TeV~\cite{atlas} (left) and $\sqrt s =13$~TeV~\cite{cms_2022} (right). }
\label{fig:mass_dep_cms_atlas}
\end{figure}
\vskip 0.5 cm
Figure~\ref{fig:mass_dep_cms_atlas} shows $q_s$ as a function of the DY pair invariant mass, obtained by comparing CASCADE3 predictions for two values of $q_0 = 1$~GeV and $q_0 = 2$~GeV with cross section measurements as functions of transverse momentum od the DY pairs produced at collision energies of 8~\cite{atlas} and 13~TeV~\cite{cms_2022}. 
The figure shows no dependence of $q_s$ on the invariant mass within the existing uncertainties.  Based on the available measurements and uncertainties from the LHC, $q_s$ remains independent of the DY pair invariant mass for $q_0 \le 2$~GeV. This suggests that the fraction of soft parton contributions with the transverse momentum  around 1~GeV  
which populate the DY $p_{\rm T}$ region up to 2-3~GeV is similar across different mass regions.   
\\

\begin{figure} [t!]
\centering
\includegraphics[width=.46\linewidth]{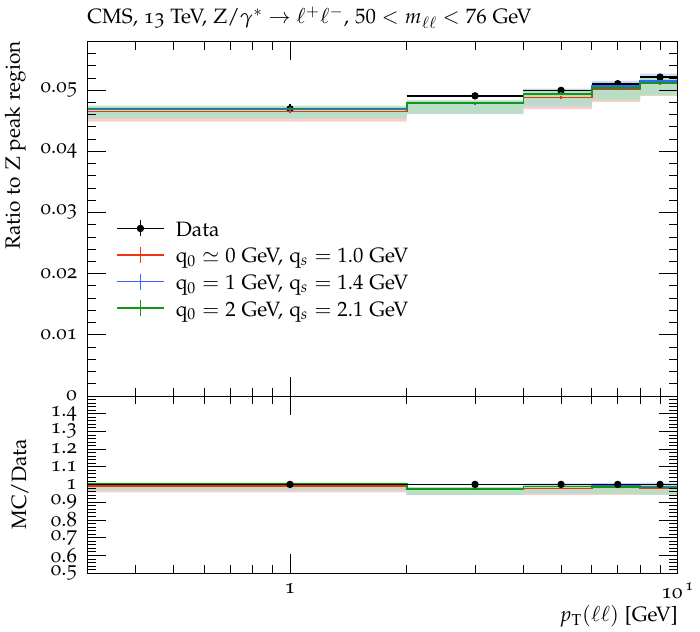}
\includegraphics[width=.46\linewidth]{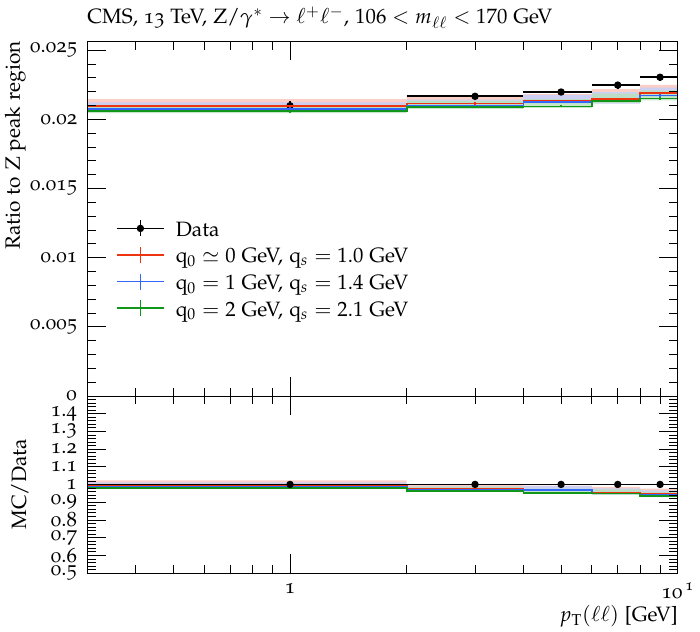}\\
\includegraphics[width=.46\linewidth]{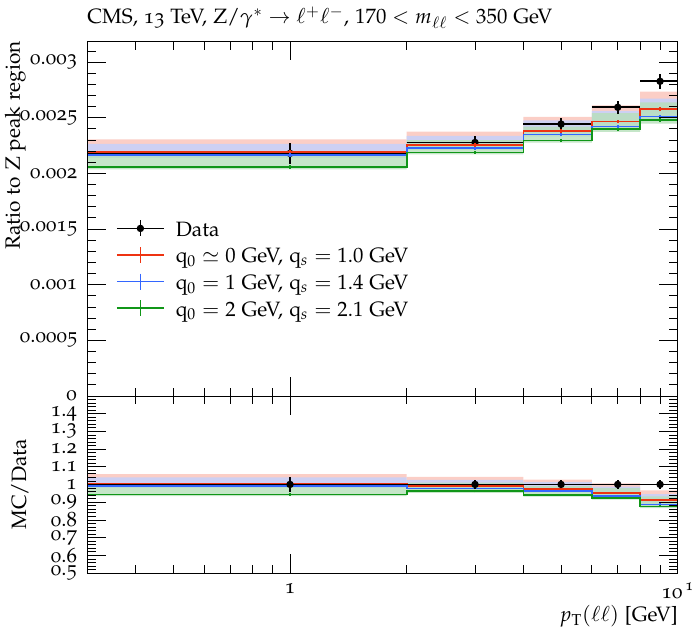}
\includegraphics[width=.46\linewidth]{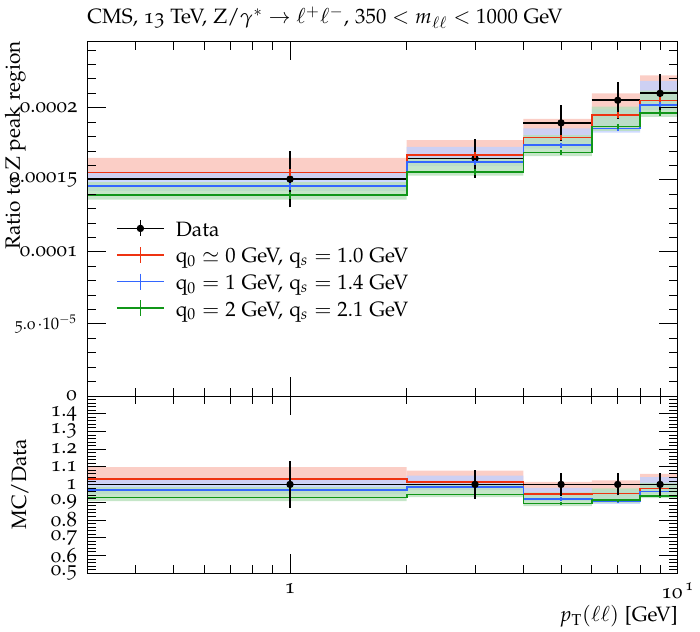}
\caption{The measured cross section ratio  to Z peak region as a function of DY transverse momentum obtained at $\sqrt s   = 13$~TeV~\cite{cms_2022}  in four mass bins  compared with CASCADE3 predictions using  three values of $q_0$: 0.01, 1 and 2 GeV. The interval of invariant mass to which the histogram refers is indicated above the histogram.
The bands show the scale uncertainty.}
\label{cms_z}
\end{figure}

Figure~\ref{cms_z} shows the ratio of the DY production cross section to the Z peak region as a function of the pair transverse momentum, measured at $\sqrt s = 13$~TeV~\cite{cms_2022}. This ratio is compared with CASCADE3 predictions obtained using three values of the cut-off parameter $q_0 = 0.01, 1, 2$~GeV, with the optimal value of $q_s$ determined in the $\chi^2 (q_s)$ minimization procedure for each $q_0$. The figure demonstrates that the ratio is well described by the predictions, which use the same $q_s$ width for each mass bin, and that the distributions obtained for different values of $q_0$ are similar.  This indicates that the relative contributions of soft gluons are quite similar across different invariant mass bins, not only in the non-perturbative region but also in the transition region ($4 \lesssim p_{\rm T}(ll) \lesssim 15$~GeV).

\begin{figure} [h!]
\centering
\includegraphics[width=.46\linewidth]{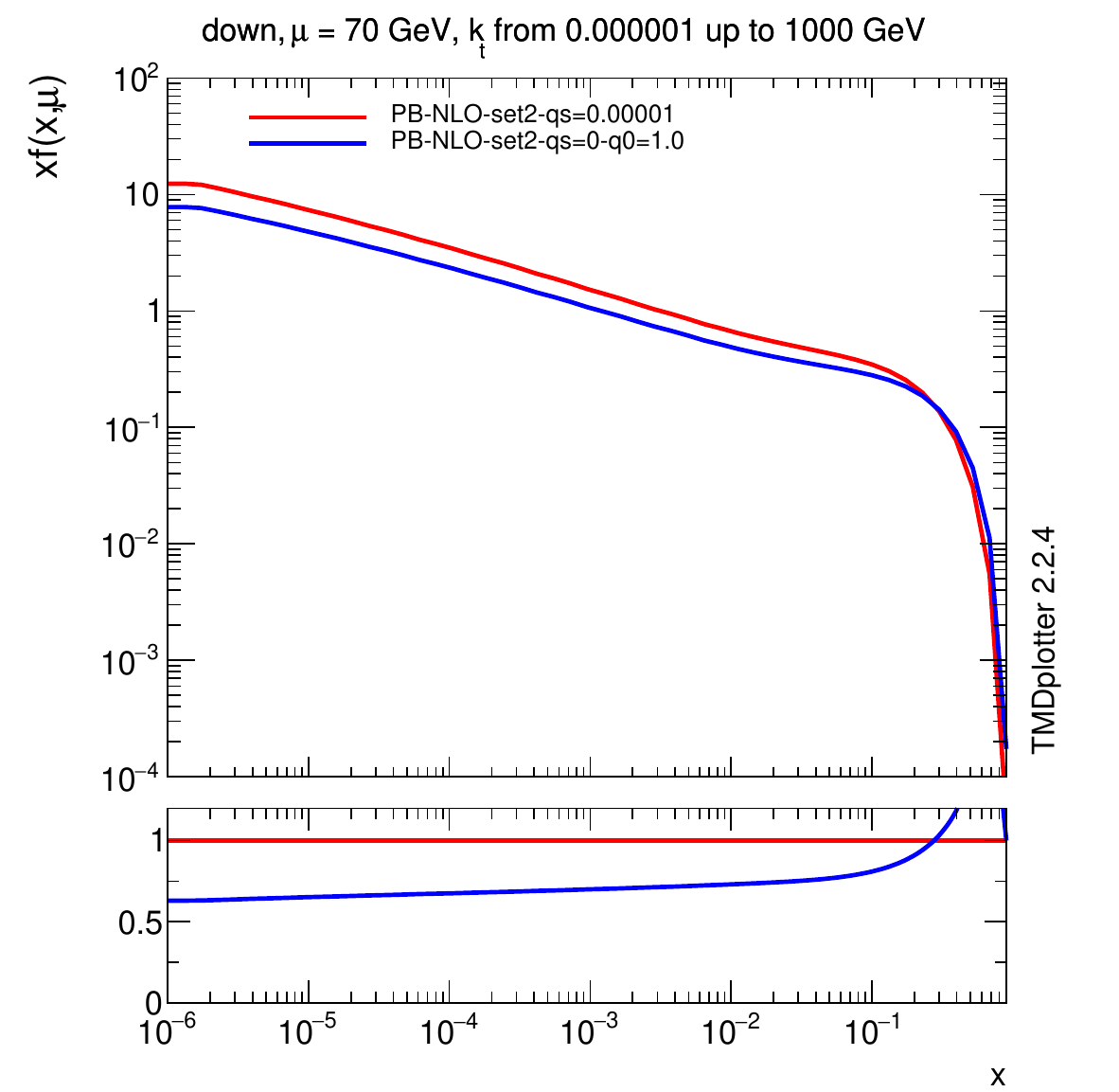}
\includegraphics[width=.46\linewidth]{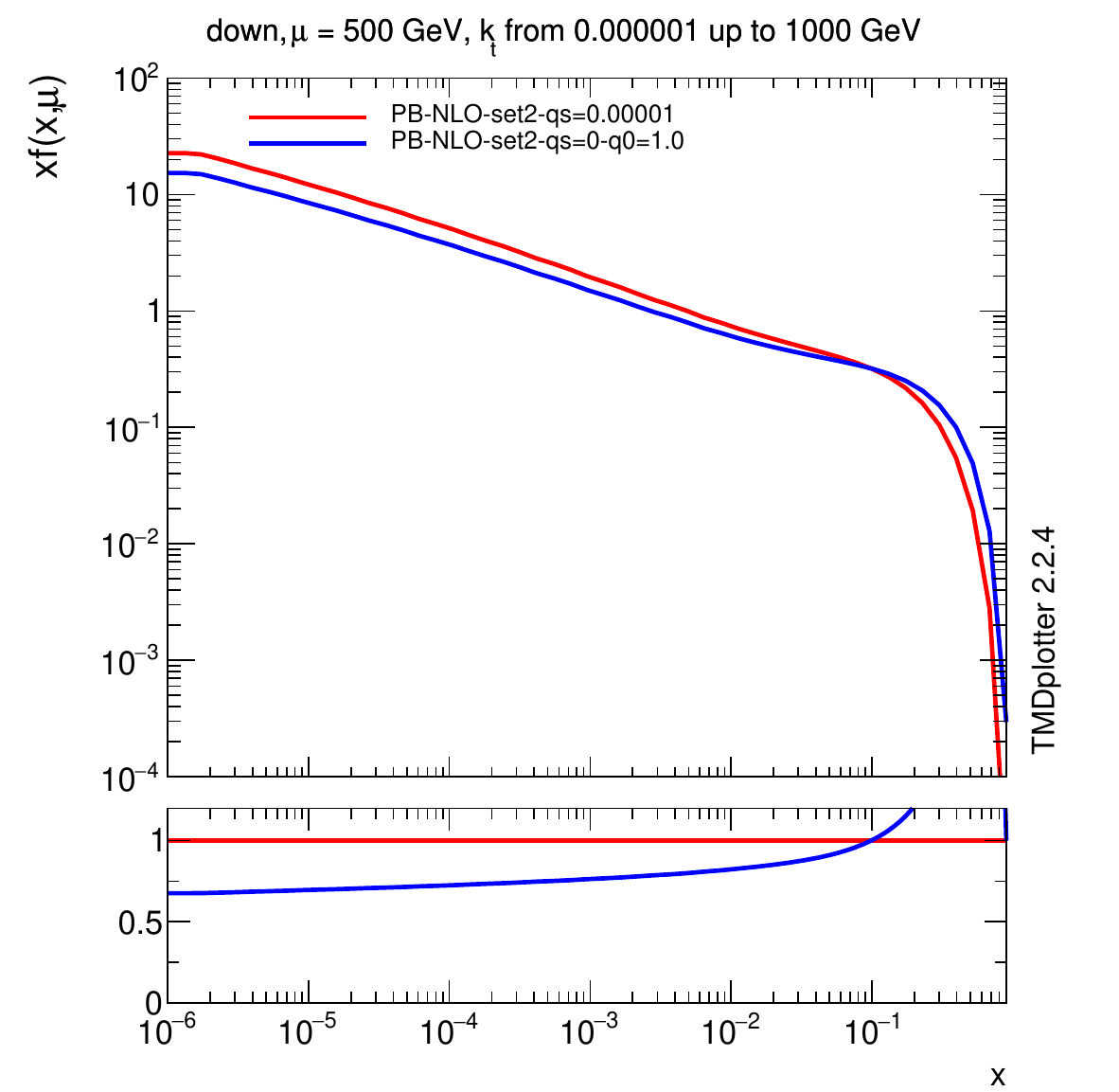}
\caption{Integrated PDFs for down quarks obtained from the PB-NLO-2018 Set2 TMD, with $q_0 \simeq 0$~GeV (red histogram) and $q_0 =1$~GeV (blue histogram) and their ratios,  for two scale values: $\mu = 70$~GeV (left) and $\mu = 400$~GeV (right). }
\label{fig:intpdf1}
\end{figure}

\vskip 0.5 cm
To examine whether the independency of the $q_s$, which reflects the independency of the soft gluon emission distribution  on  the DY pair invariant mass, is a consequence of the insensitivity of the kinematic range of the measurements to the scale $\mu$ in the Sudakov form factor, we examine how the integrated PDF changes when applying a $q_0$ cut.  Figure~\ref{fig:intpdf1} shows the PDFs obtained by integrating the down quark TMD set referred to as as PB TMD Set2~\cite{pb,updf} over parton transverse momenta, $k_{\rm T}$, for the two values of $q_0$: $q_0 \simeq 0$~GeV and $q_0=1$~GeV for the scale values $\mu$ of 100 and 500~GeV. The PDF plots are obtained from the graphical interface TMDplotter~\cite{tmdplotter1}. This figure demonstrates that the change in the integrated PDFs upon introducing a cut of $q_0 \sim 1$~GeV is significant, but similar across a wide range of scale values $\mu$ relevant to the measurements available at the LHC. This is consistent with the result of $q_s$ vs DY pair invariant mass obtained from LHC measurements over a wide range of pair invariant masses. Furthermore, it aligns with the non-perturbative Sudakov form factor, which is sensitive at small values  of $\mu$  and changes slowly in the wide range of $\mu$, from 100 to 500~GeV, where the LHC measurements have been performed. 

\begin{figure} [h!]
\centering
\includegraphics[width=.46\linewidth]{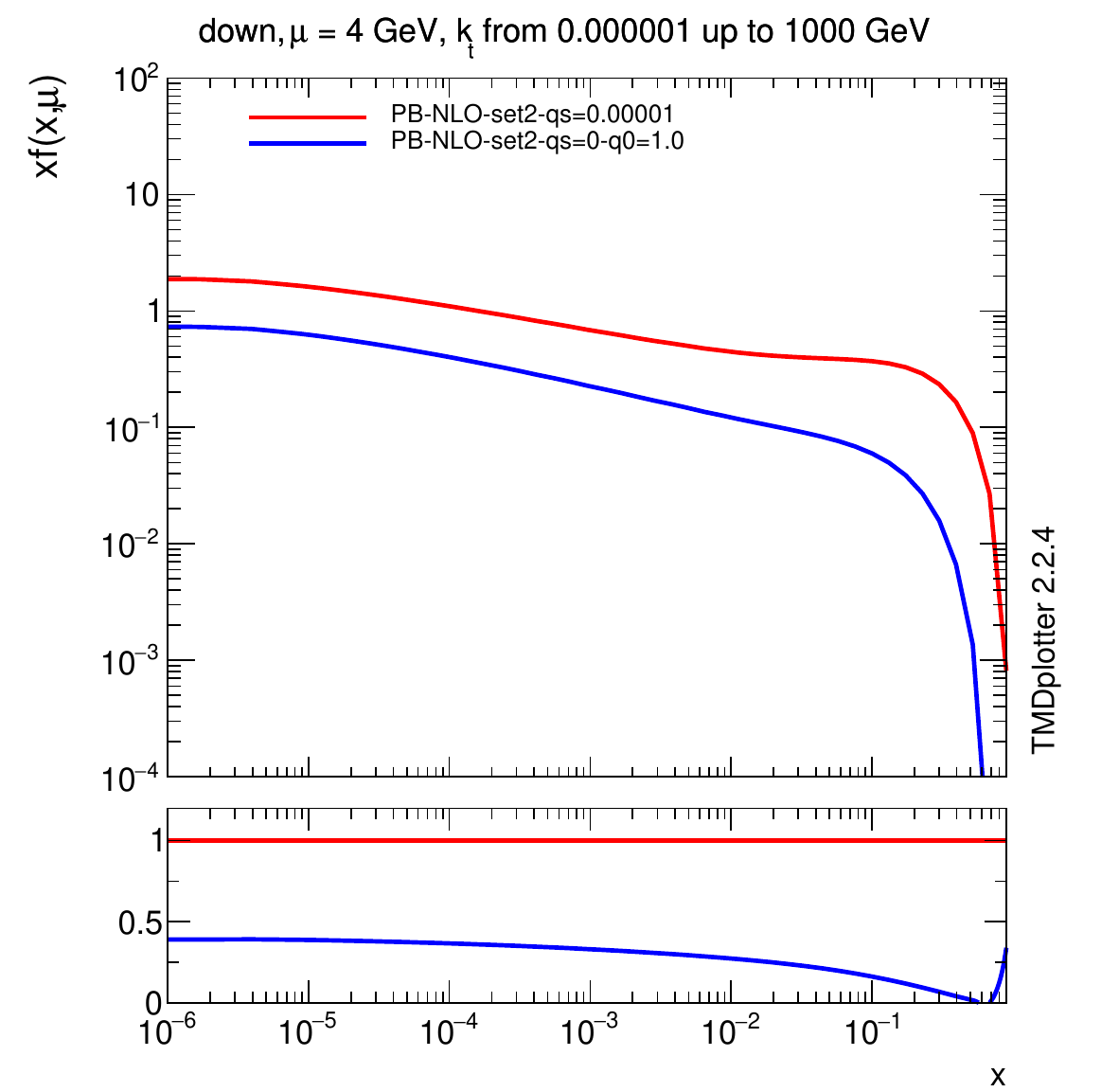}
\includegraphics[width=.46\linewidth]{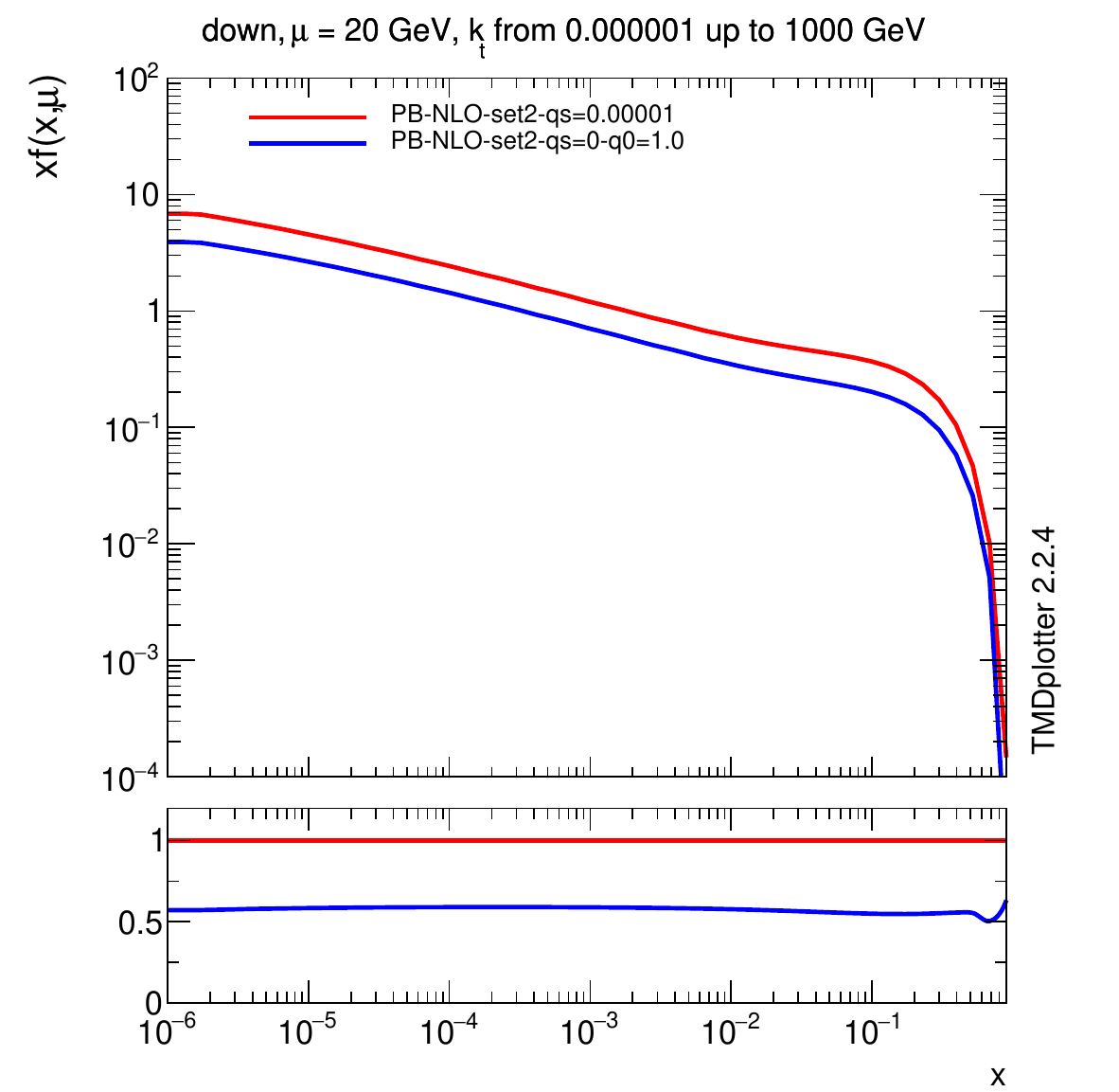}
\caption{Integrated PDFs for down quarks obtained from the PB-NLO-2018 Set2 TMD, with $q_0 \simeq 0$~GeV (red histogram) and $q_0 =1$~GeV (blue histogram) and their ratios,  for two scale values: $\mu = 70$~GeV (left) and $\mu = 400$~GeV (right). }
\label{fig:intpdf2}
\end{figure}

\vskip 0.5cm
Figure~\ref{fig:intpdf2} shows how the integrated PDF for down quark changes with the inclusion of the $q_0$ cut when the $\mu$ value is much smaller than those corresponding to the measurements from the LHC. The figure displays the integrated PDFs without and with the cut of $q_0 = 1$~GeV for  two values of $\mu = 4$~GeV and  $\mu = 20$~GeV  which cover a much smaller range in $\mu$ than the values in the previous figure. In the case of a small scale value, the figure shows a more pronounced difference in the change of the PDFs for the two values of $\mu$, indicating that the change in the integrated PDFs due to the $q_0 \sim 1$~GeV cut varies rapidly with the $\mu$ scale, which is relevant for DY pair masses of several GeVs. There is a  different trend at low $\mu$, which corresponds to measurements of pair $p_{\rm T}$ at low invariant masses not yet available at the LHC.  From this, we can conclude that the relative amount of soft gluons removed by the cut changes significantly at low scales, and measurable changes in the value of $q_s$ could be expected at low DY pair invariant masses.

\begin{figure} [h!]
\centering
\includegraphics[width=.45\linewidth]{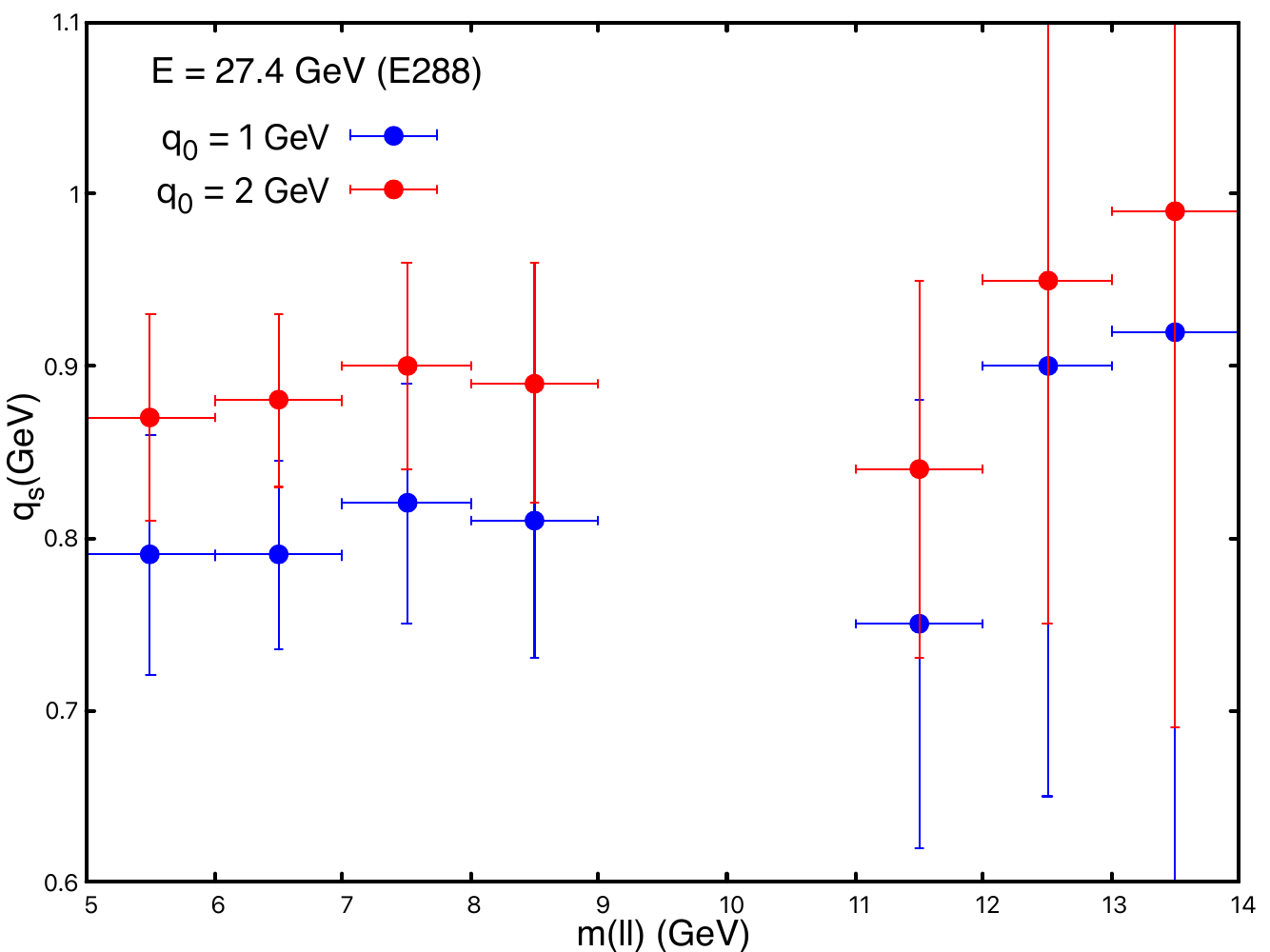}
\includegraphics[width=.45\linewidth]{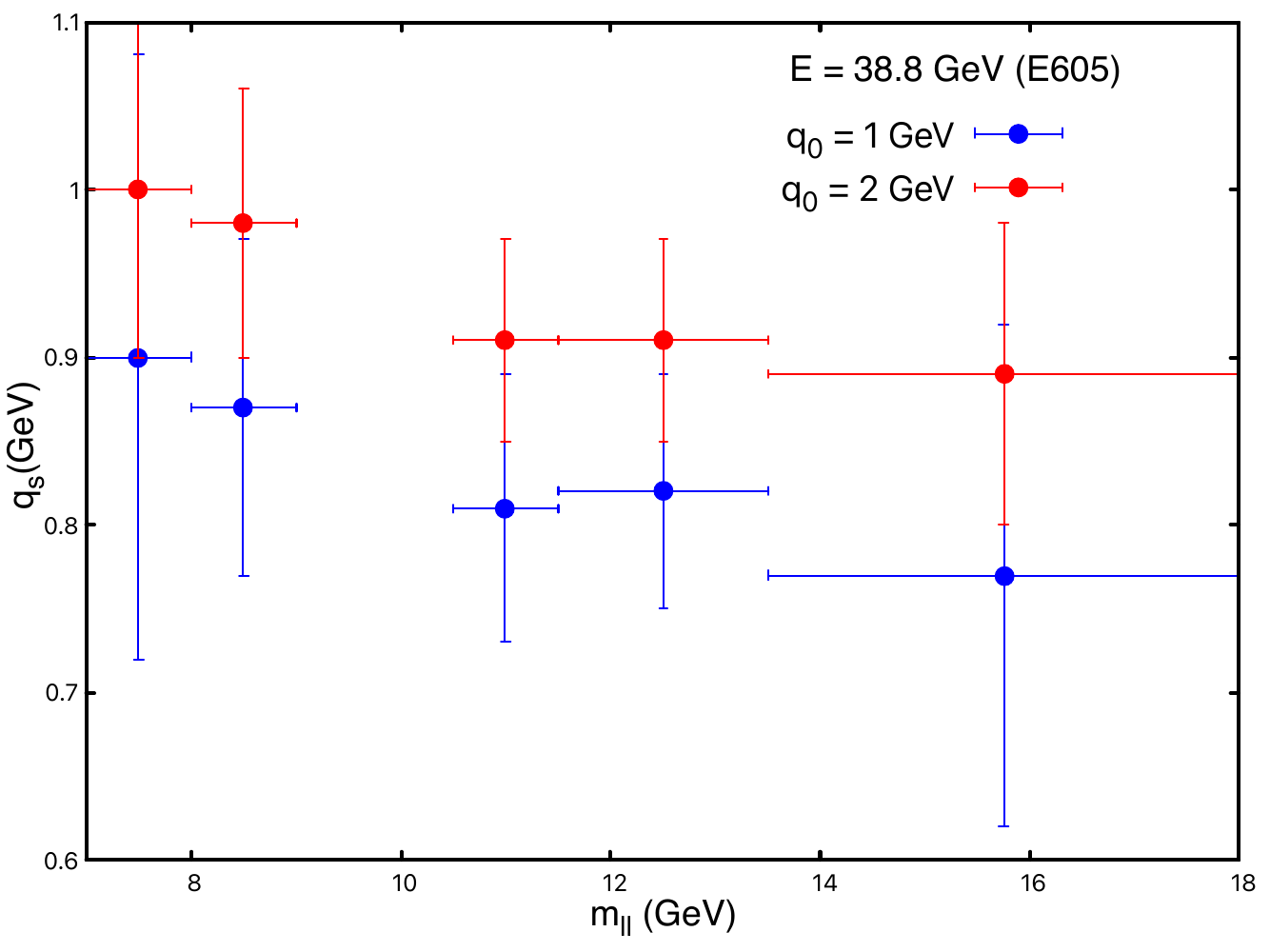}
\caption{Intrinsic $k_{\rm T}$ width as a function of DY pair invariant mass, obtained from the comparison of CASCADE 3 predictions with two values of $q_0$ and measurements obtained at $\sqrt s = 27.4$~GeV~\cite{e288} (left) and $\sqrt s =38.8$~GeV~\cite{e605} (right).}
\label{fig:mass_dep_e288_e605}
\end{figure}

To investigate the region of very small DY pair invariant masses, we have determined the mass dependence of $q_s$  using available measurement obtained at $\sqrt s = 27.4$~GeV~\cite{e288}, and at $\sqrt s = 38.8$~GeV~\cite{e605}.  This is shown in Figure~\ref{fig:mass_dep_e288_e605}.

This figure shows a weak dependence of $q_s$ and the soft gluon contribution on the pair invariant mass. To achieve masses on the order of several GeV at these collision energies,  a large value of the longitudinal momentum fraction carried by a colliding parton, $x$, is required since $M^2_{{\rm {DY}}} = x_1 \cdot x_2 \cdot s$. 
Therefore, at this collision energies, DY pairs with low invariant masses  are primarily produced at high $x$ from valence contributions, making it unlikely to expect a measurable dependence of $q_s$ on DY pair invariant mass. Although the errors are large and it is not possible to draw a firm conclusion, the  trend of $q_s$ changing  with $m{(ll)}$ is noticeable in the measurements obtained  at $\sqrt s = 38.8$~GeV. 

\section{Influence of QED radiation on very low transverse momentum distributions of DY pairs}
\label{sec:qed}

\begin{figure} [h!]
\includegraphics[width=.33\linewidth]{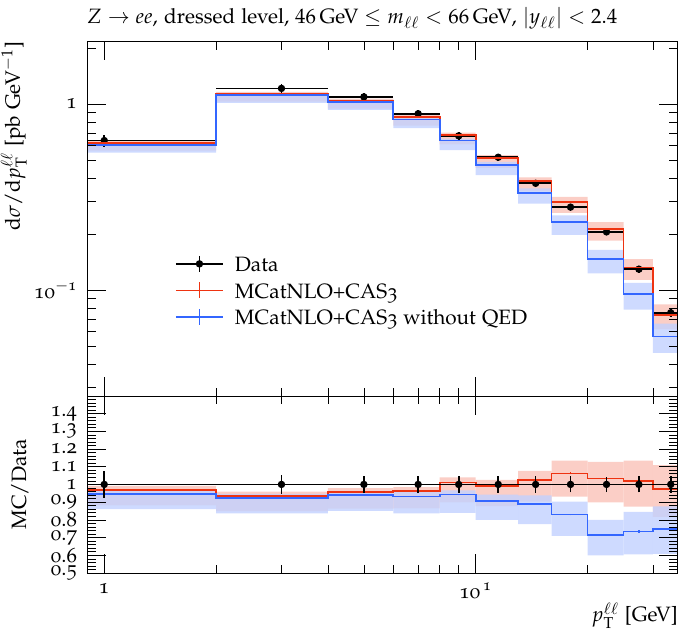}
\includegraphics[width=.33\linewidth]{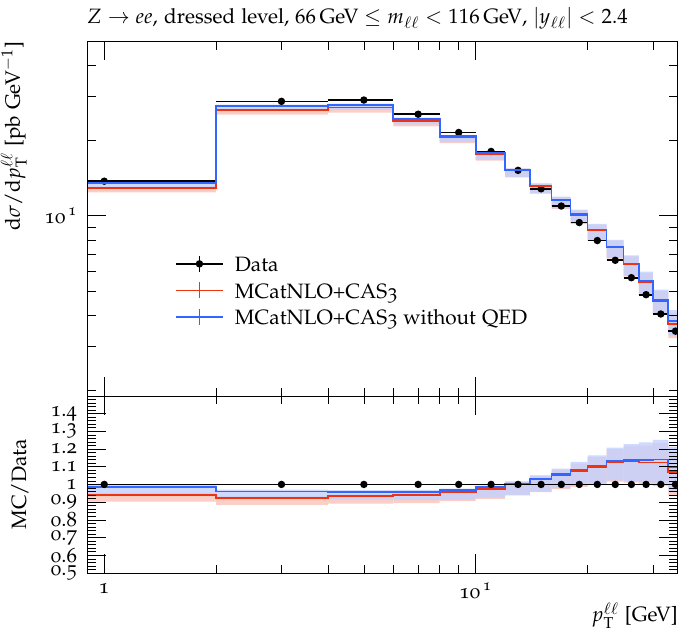}
\includegraphics[width=.33\linewidth]{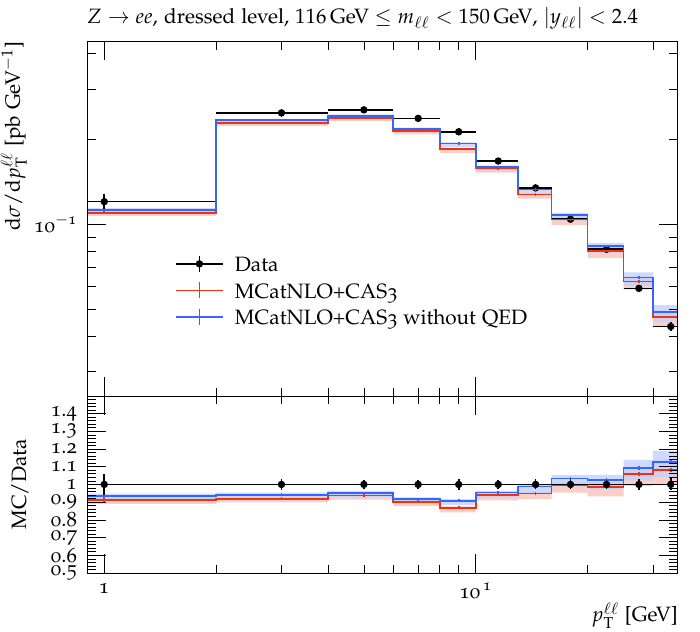}
\caption{CASCADE3 predictions with (red histogram) and without (blue histogram)  final state QED radiation,  along with the measured cross section as a function of the transverse momentum of DY pairs produced at $\sqrt s = 8$~TeV~\cite{atlas}, for three mass bins indicated in the histograms: $46 \;  {\rm {GeV}} \le m(ll) \le  \; 66 \; {\rm {GeV}}$ (left), 
$66 \; {\rm {GeV}} \le m(ll) \le  116 \; {\rm {GeV}}$ (middle) and $116 \; {\rm {GeV}} \le m(ll) \le  150 \; {\rm {GeV}}$ (right). The bands show the scale uncertainty.}
\label{fig:atlas_qed}
\end{figure}

The impact of final state QED radiation on the transverse momentum distributions of DY pairs created in proton-proton interactions at 13 TeV~\cite{cms_2022} was studied in  paper~\cite{ktpaper}. It was shown that QED radiation has a significant effect on the transverse momentum distribution for invariant masses bellow Z peak region, down to about 30~GeV. Very similar effect is observed at the somewhat smaller collision energy of 8 TeV~\cite{atlas} as shown in figure~\ref{fig:atlas_qed}. This figure compares data  with CASCADE3 predictions, both with and without final state QED radiation, for the three mass bins in the range from 46 to 150 GeV. 

Although there is a significant impact from QED radiation below the Z-peak region, it does not affect the determination of the intrinsic-$k_{\rm T}$ width and does not interplay with the most relevant non-perturbative contributions that populate the lowest DY pair transverse momentum region as the effect becomes noticeable for $p_{\rm T} (ll) \gtrsim 8$~GeV. 
\\

\begin{figure} [h!]
\includegraphics[width=.33\linewidth]{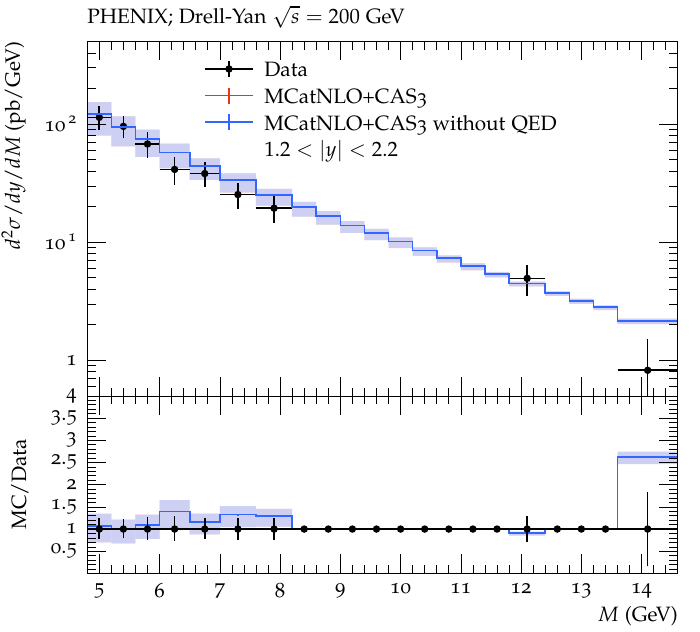}
\includegraphics[width=.33\linewidth]{ 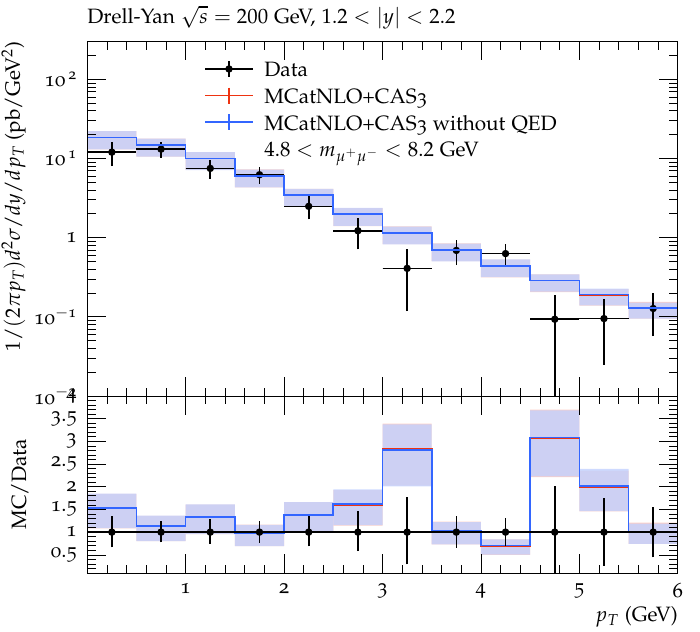}
\includegraphics[width=.33\linewidth]{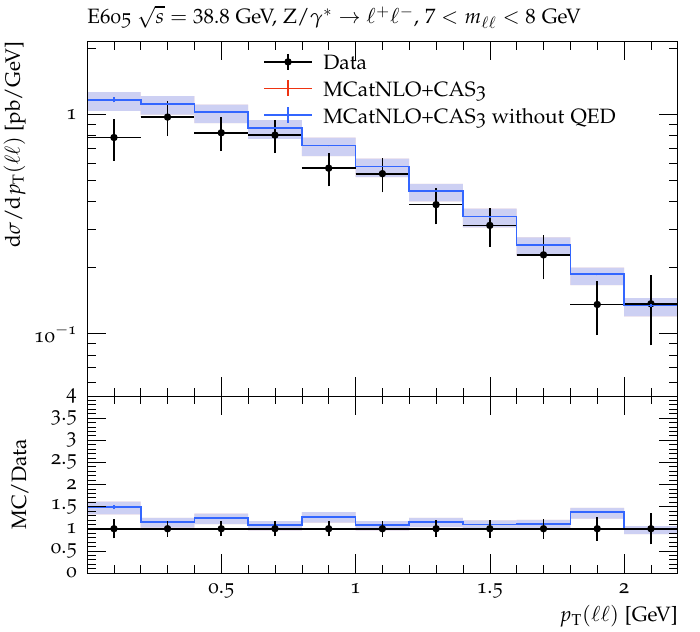}
\caption{Comparisson of CASCADE3 predictions with (red histogram) and without (blue histogram)  final state QED radiation with: the measured cross section as a function of DY pair invariant mass obtained  at $\sqrt s   = 200$~GeV~\cite{phenix} (left), the measured cross section as a function of DY pair  transverse momentum obtained at $\sqrt s   = 200$~GeV~\cite{phenix} (middle) and at $\sqrt s   = 38.8$~GeV~\cite{e605} (right). The bands show the scale uncertainty.}
\label{fig:lowmass_qed}
\end{figure}

An impact of QED radiation at  low pair invariant masses used to study internal transverse motion and soft gluon emission as well as their interplay was also examined. Figure~\ref{fig:lowmass_qed} shows invariant mass and transverse momentum distributions of DY pairs measured at $\sqrt s = 200$~GeV~\cite{phenix} as well as transverse momentum distribution in $4.2 < m(ll) < 8$~GeV measured at 38.8~GeV~\cite{e605}. The measurements in this figure are compared with CASCADE3 predictions with and without inclusion of QED radiation. As this figure shows, there is no impact of the QED radiation at low invariant mass of the order of several GeV and the transverse momentum distributions are identical. 

\begin{figure} [h!]
\centering
\includegraphics[width=.42\linewidth]{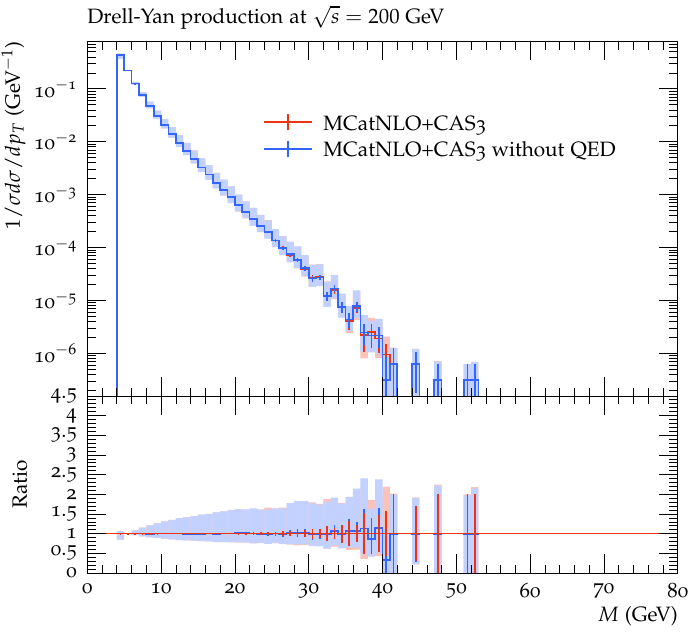}
\includegraphics[width=.42\linewidth]{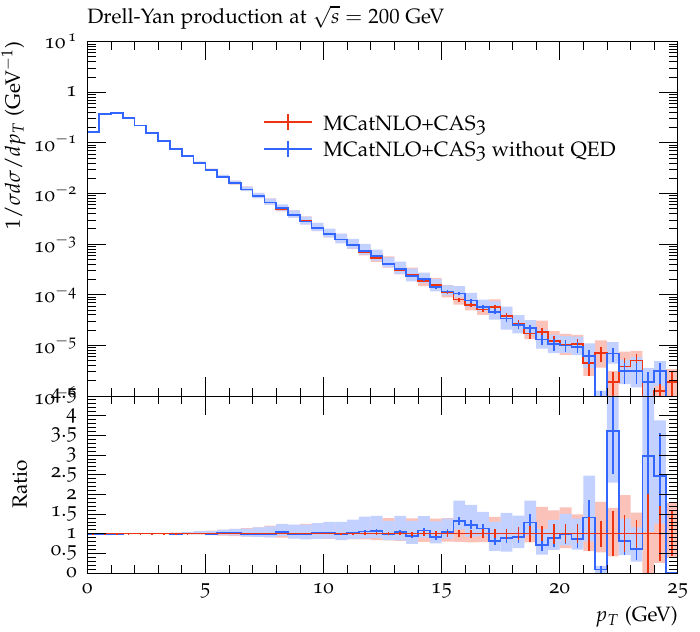}
\caption{ CASCADE3 predictions of the cross section as a function of DY pair invariant mass (left) and pair transverse momentum (right) at $\sqrt s   = 200$~GeV  with (red histogram) and without (blue histogram)  final state QED radiation in the wide range of pair invariant masses and transverse momenta. The bands show the scale uncertainty.}
\label{fig:phenix_extended}
\end{figure}

To examine QED radiation at lower center of mass energy in more detail, we studied the DY pair invariant mass and transverse momentum distributions, extending the range of available measurements at $\sqrt s = 200$~GeV. Figure~\ref{fig:phenix_extended} shows CASCADE3 predictions for invariant mass and transverse momentum distributions of DY pairs, both with and without QED radiation. As seen in the figure, the leptons also radiate at low $\sqrt s$, but this effect is not pronounced due to the shape of the DY invariant mass distribution. When a  lepton radiates around the Z-mass, there is a significant effect, as the mass will be lower, leading to a migration from the Z peak to lower DY masses. At much lower (and higher) DY masses,  QED radiation still occurs, but it arises from a continuous mass spectrum. The migration from higher to lower DY masses is negligible because of the falling DY invariant mass spectrum. The figure indicates that the radiating leptons contribute to pairs with larger transverse momentum. 

\section{Conclusion}
\label{sec:concl}

The study presented in this paper provides a detailed explanation of the influence of the Sudakov form factor on the determination of the parton intrinsic-$k_{\rm T}$ distribution using
the PB Method.  Previous studies indicated that the energy dependence of intrinsic-$k_{\rm T}$, observed in shower-based event generators, is related to the Sudakov form factor, whose non-perturbative part is neglected through the integral over the $z$ resolution scale. The impact of the Sudakov form factor on the intrinsic-$k_{\rm T}$ width is associated with the soft gluon contribution and its interplay with the contribution from the parton internal motion at the lowest transverse momenta of the DY pairs, where two non-perturbative processes dominate.

To confirm the impact of the Sudakov form factor and study the relative contribution of soft gluons, we conducted a detailed analysis of the Sudakov form factor's influence through its dependence on the evolution scale $\mu$. For this, we used the CASCADE3 event generator, which is based on the PB Method. In this generator, the shower-based Monte Carlo event generation was mimicked by imposing a requirement for a minimum parton transverse momentum emitted at the branching, $q_0$. 
 The study examines  the dependence of the determined intrinsic-$k_{\rm T}$ width on the invariant mass of the DY pair, which is directly related to the scale $\mu$ in the Sudakov form factor. Although we did not confirm such dependence, we indicated that it is related to the insensitivity of the available measurements to the scale, which we demonstrated by analysing the relative change of the integrated PDF while applying the $q_0$ cut. It was shown that the relative change of the PDFs is very similar over a wide range of $\mu$ relevant for the available measurements.

Therefore, given the available measurements and existing uncertainties, we cannot confirm the dependency of the relative soft gluon contribution on the DY pair invariant mass which is  consistent with the behaviour (i.e. the small sensitivity) of the non-perturbative Sudakov form factor in the measured regions.

Part of if this study was dedicated to examining the impact of final state QED  radiation on the non-perturbative processes discussed here. Final state QED radiation has a significant effect in the invariant mass region below the Z-peak,  influencing the transverse momentum distributions of DY pairs not only quantitatively, but  also in terms of shape. 
Although there is a significant impact of final-state QED radiation below the Z-peak region, it becomes relevant only for $p_{\rm} T(ll) \gtrsim 8$~GeV, and therefore does not overlap with the contribution from non-perturbative processes. 
Regarding the available measurements in the region of small DY pair invariant masses and small transverse momentum at low collision energies, QED radiation does not have an impact; it becomes relevant only at large values of the transverse momentum of DY pairs. Therefore, it was concluded that QED radiation has no effect on the determination of the intrinsic-$k_{\rm T}$ width and does not interplay with the main non-perturbative processes studied here. 
\\

The next step will be a detailed study and confirmation of the effects related to the interplay between parton internal motion and soft gluon emission, focusing on the impact of the Sudakov form factor, using a shower-based Monte Carlo generator directly.

 \vskip 0.5 cm 
\noindent 
{\bf Acknowledgments.} 
 I am grateful to the organisers of the XIII International Conference on New Frontiers in Physics (ICNFP 2024) for the opportunity to give a  talk on the presented results which are based on the collaborative work in the CASCADE group.  I wish to acknowledge all the colleagues for the very fruitful and extensive collaboration. 

 \vskip 0.5 cm 
\noindent 
{\bf Funding.} 
The results presented here are part of a national scientific project that has received
 funding from Montenegrin Ministry of Education, Science and Inovation.

\end{document}